\documentclass{iau} 
\usepackage{graphicx}

\title[JD 11.~~Variability during the AGB - post-AGB transition] 
{The loss of large amplitude pulsations at the end of AGB evolution}

\author[Dieter Engels et al.]   
{D.  Engels$^1$,
S. Etoka$^2$, \and E. G\'{e}rard$^3$}

\affiliation{$^1$Hamburger Sternwarte, Universit\"at Hamburg, Germany, email: {\tt dengels@hs.uni-hamburg.de}\\[\affilskip]
$^2$Jodrell Bank Centre for Astrophysics, University of Manchester, UK, email: {\tt sandra.etoka@googlemail.com}\\[\affilskip]
$^3$GEPI, Observatoire de Paris, Meudon, France, email: {\tt eric.gerard@obspm.fr}}

\pubyear{2018}
\volume{343}  
\jname{Why Galaxies Care about AGB Stars}
\editors{F. Kerschbaum, M. Groenewegen \& H. Olofsson, eds.}
\begin{document}

\maketitle

\begin{abstract}
Since 2013 we are performing with the Nancay Radio Telescope (NRT) a
monitoring program of $>100$ Galactic disk OH/IR stars, having bright
1612-MHz OH maser emission. The variations of the maser emission are
used to probe the underlying stellar variability. We wish to
understand how the large-amplitude variations are lost during the AGB
-- post-AGB transition.  The fading out of pulsations with steadily
declining amplitudes seems to be a viable process.  
\end{abstract}

Stars evolving on the thermal-pulsing Asymptotic Giant Branch (AGB)
are in general observed as large-amplitude variables, but are almost
non-variable in the post-AGB phase.  In models covering the AGB
--post-AGB transition, the evolutionary timescales depend on the
assumptions of the change of the mass-loss rates. They must drop on
short timescales from late AGB values of $10^{-5} - 10^{-4}$ to
post-AGB values of $10^{-7} - 10^{-8}$ $M_{\odot}$\,yr$^{-1}$.  While
the mass loss rates are parametrized on the AGB as a function of
pulsation period, they are completely unconstrained starting with the
time after which the pulsation ceased until the time that a radiation
driven wind as observed in Planetary Nebulae takes over (\cite[Miller
  Bertolami 2016]{Miller16}; MB16 hereafter).

Towards the end of AGB evolution, stars can develop very high
mass loss rates, which enshrouds them completely by dust and
gas. Among them are the OH/IR stars, which encompass large-amplitude
variables on the AGB (L-AGB stars) with periods $\sim700-2000$ days
and almost non-variable stars (S-pAGB: small amplitude post-AGB stars,
including 'non-variable' stars), which are thought to evolve in the
early post-AGB phase. In both phases, the stars are still deeply
embedded in their dusty circumstellar shell. H$_2$O and OH maser
emissions are present in both phases.
The association of the S-pAGB
stars with the post-AGB phase is supported by observations that some
of them already have diluted dust shells (\cite[Engels
  2007]{Engels07}), which indicate a recent decrease of the mass loss
rates, and that others show prominent bipolar outflows
(f.e. OH17.7--2.0 = IRAS 18276--1431, \cite[S\'{a}nchez-Contreras et
  al. 2007]{Sanchez07}; OH 53.6--0.2 = IRAS 19292+1806, \cite[Sahai et
  al. 2007]{Sahai07}) including ``water fountains'' (f.e. W43A = OH
31.0+0.0 = IRAS 18450--0148, \cite[Chong et al. 2015]{Chong15}). It
is during the obscured phase that (at least in the more massive stars)
the AGB -- post-AGB evolutionary transition takes place and the stars
stop pulsating.

Monitoring the stars via their bright and relatively stable OH maser
emission is needed, because especially the S-pAGB candidates have very
red spectral energy distributions, and cannot be monitored in the
optical or the near-infrared. As a basic sample to study the
transformation of the variability characteristics, we use the full
sample of OH/IR stars of \cite[Baud et al. (1981)]{Baud81}, updated by
\cite[Engels \& Jim\'{e}nez-Esteban (2007)]{Engels07a}. This "Bright
OH/IR star sample" comprises 115 stars, with almost all located at
$10<l<150^\circ$, $\mid$b$\mid$ $< 4^\circ$ along the Galactic
plane. It is quite complete for bright 1612-MHz OH
masers ($F_\nu > 4$ Jy). The brighter part of the sample has been
monitored by \cite[Herman \& Habing (1985)]{Herman85}(herafter HH85),
who reported several sub-groups with different amplitudes and
periodicity among S-pAGB stars. Objects, which are currently
transiting from L-AGB to S-pAGB variability may hide in the sample. To
find them, we are monitoring since 2013 the 1612-MHz OH masers with
the NRT, to probe the underlying stellar variability.

In our sample, the L-AGB and S-pAGB stars are almost of equal number.
Assuming similar OH maser luminosities, this implies that the
``pulsating'' phase connected to relatively high mass-loss rates
($\dot{M} > 10^{-5}$ M$_{\odot}$/yr) is of similar duration as the
early post-AGB phase (\cite[Engels 2002]{Engels02}). OH/IR stars must
have experienced hot bottom burning on the AGB to avoid being
converted to carbon-rich stars, and as such they must have had massive
progenitors on the main sequence $M\ge3$ $M_{\odot}$. According to
MB16, the predicted transition times $\tau_{tr}$ during the early (and
obscured) post-AGB phase until the optical reappearance of the central
stars last only $<1000$ years. In the later post-AGB phase the dust
shells are dispersed, and, in general, maser emission
disappear. Assuming a minimum lifetime of the OH maser emission in
the ``Bright OH/IR stars sample'' of 2000 years (\cite[Engels \&
  Jim\'{e}nez-Esteban 2007]{Engels07b}), the time massive AGB stars appear
as obscured OH/IR stars can last only a few thousand years.

As of May 2018, we have the variability characterizations for 52 stars
(34 L-AGB, 18 S-pAGB). Another 28 stars are currently (2018/2019)
monitored to obtain a characterization, while the remaining stars are
planned to be monitored in 2020/2021. Monitoring of newly recognized
L-AGB stars is continued until the period is determined.  S-pAGB stars
are re-observed occasionally to search for long-term trends, such as
found by Wolak et al. (2014). They reported that the OH maser of one
of the S-pAGB stars, OH 17.7--2.0, is continuously fading since its
discovery and predict that the maser will fall below the detection
limit around 2030.  While the L-AGB stars in HH85 are confirmed, some
of their S-pAGB stars had to be reclassified as L-AGB variables with
periods $P>1000$ days.  Among 15 OH/IR stars not monitored by HH85, we
found 9 L-AGB stars (60\%), while the reminder shows at most irregular
fluctuations qualifying them as S-pAGB stars.

No stars with short-period, small amplitude pulsations have been
found, as assumed to exist as transition objects by \cite[Bl\"ocker
  (1995)]{Bloecker95}. However, we found a couple of stars, which show
periodic variations with periods similar to those of L-AGB stars but
with significantly smaller amplitude (\cite[Engels et
  al. 2018]{Engels18}). We consider them as the best candidates for
transition objects. While an instantaneous cessation of the pulsation
(\cite[Vassiliadis \& Wood 1994]{Vassiliadis94}; MB16) cannot be ruled
out, we consider the fading out of pulsations with steadily declining
amplitudes (damped oscillator) as a viable process.

\end{document}